\documentclass[final]{IEEEtran}
\usepackage[pdftex]{graphicx}
\usepackage{epstopdf}
\makeatletter
\def\ps@headings{%
\def\@oddhead{\mbox{}\scriptsize\rightmark \hfil \thepage}%
\def\@evenhead{\scriptsize\thepage \hfil \leftmark\mbox{}}%
\def\@oddfoot{}%
\def\@evenfoot{}}
\makeatother
\usepackage[cmex10]{amsmath}
\usepackage[pdftex]{graphicx}
\usepackage{epstopdf}
\usepackage{cite,setspace,url}
\usepackage{latexsym}
\usepackage{verbatim}

\usepackage[cmex10]{amsmath}
\usepackage{amssymb}

\newtheorem{theorem}{Theorem}

\usepackage{mathtools}
\usepackage{texdef2015}
\usepackage[normalem]{ulem}
\usepackage[makeroom]{cancel}
\allowdisplaybreaks 

\newcommand{\age}{\Delta}

\newlength{\swwidth}

\newcommand{\vbset}[1]{\vbar_{\set{#1}}}
\newcommand{\neighbor}[1]{\lambda_{#1}}
\newcommand{\Ncal}{\mathcal{N}}

\newcommand{\Lcal}{\mathcal{L}}
\newcommand{\Qcal}{\mathcal{Q}}

\newcommand{\R}{\mathbb{R}}

\newcommand{\qmax}{q_{\max}}

\newcommand{\zerov}[1][\mbox{}]{\mathbf{0}_{#1}}

\renewcommand{\vec}[1]{\begin{bmatrix} #1\end{bmatrix}}

\newcommand{\nlen}{n} 

\usepackage{tikz}
\usetikzlibrary{automata,arrows,positioning,calc,shapes.multipart,chains,shapes}

\usepackage{ifthen}

\newcommand{\Others}[1]{\ifthenelse{\equal{#1}{1}}{2}{1}}

\newcounter{lettercount}

\begin{document}

\newboolean{mytwocolumn}
\setboolean{mytwocolumn}{false}

\title{The Age of Gossip in Networks}
\author{\IEEEauthorblockN{Roy D.~Yates\thanks{This work was supported by NSF award CCF-1717041.}
}\\
\IEEEauthorblockA{WINLAB, Department of Electrical and Computer Engineering\\
Rutgers University\\
{\em ryates@winlab.rutgers.edu}}
}

\maketitle
%
\begin{abstract}
A source node updates its status as a point process and also forwards its updates to a network of observer nodes.  Within the network of observers, these updates are forwarded as point processes from node to node. Each node wishes its knowledge of the source to be as timely as possible. In this network, timeliness is measured by a discrete form of age of information:  each status change at the source is referred to as a version and the age at a node is how many versions out of date is its most recent update from the source. This work introduces a method for evaluating the average version age at each node in the network when nodes forward updates using a memoryless gossip protocol. This method is then demonstrated by version age analysis for a collection of simple networks.  For  gossip on a complete graph with symmetric updating rates, it is shown that each node has average age that grows as the logarithm of the network size.
\end{abstract}   


\section{Introduction}\label{sec:intro}
Gossip is a popular mechanism to  convey status information in a distributed systems and  networks. The efficacy of gossip mechanisms for distributed computation  \cite{boyd2004analysis,boyd2006randomized} and message dissemination \cite{shah2009gossip}  is well known.  While it is also known that gossip mechanisms can be inefficient relative to more complex  or application-specific algorithms, it is recognized that gossip remains an attractive option in settings when protocols need to be simple or the  network topology or connectivity is time-varying \cite{birman2007promise}.  For example, gossip protocols  could be a good choice for low latency vehicular safety messaging. And yet, while vehicular message exchange was the early motivation for age of information (AoI) research \cite{Kaul-GRK-secon2011,Kaul-YG-globecom2011piggybacking}, there has been little (if any) effort to examine AoI for gossip protocols.
 
In this work, we begin to re-examine gossip from an age-of-information (AoI)  perspective \cite{Kaul-YG-infocom2012,Yates-2020survey}. Specifically, a source wishes to share its status update messages with a network of $n$ nodes. These nodes, which  can be viewed monitors of the source, employ gossip to randomly forward these update messages amongst themselves in order that all nodes have timely knowledge of the state of the source.   

This work extends AoI analysis  in a class  of  {status sampling} networks, a networking paradigm that is consistent with gossip models in that short messages, representing samples of a node's status update process, are delivered as point processes to neighbor nodes.  This  ``zero service time'' model  may be useful in a high speed network in which updates represent small amounts of protocol information (requiring negligible time for transmission)  that are given priority over data traffic. This model has also been widely used in the age analysis of enegry harvesting updaters \cite{jing-age-online, jing-age-erasures-infinite-jour, arafa-age-online-finite, arafa-age-erasure-no-fb, arafa-age-erasure-fb,elif_age_eh,liu-age-eh-sensing} where updating rates are constrained by energy rather than bandwidth.  While the transmission of a single update may be negligible, the update rates are limited so that  protocol information in the aggregate does not consume an excessive fraction of network resources. 

Prior work on status sampling networks \cite{Yates-aoi2018,Yates-it2020}  developed tools for analyzing age in line networks in which each node $i$ only received updates from node $i-1$.  The key advance of this work is the development of an average age analysis method for monitors that receive updates via multiple network paths. 

\section{System Model and Summary of Results}\label{Sec:model}
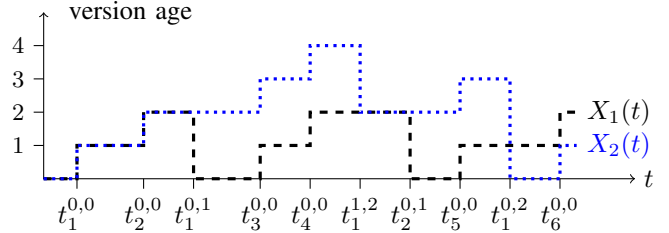
\begin{figure}[t]
\centering
\begin{tikzpicture}[baseline=(current bounding box.center),scale=\linewidth/20cm]
\draw [<->] (0,5) node [above,right] {\begin{tabular}{c} version age\end{tabular}} -- (0,0) -- (17.75,0) node [right] {$t$};
\draw (1,0) -- (1,-0.3) node [below] {$t^{0,0}_1$} 
(3,0) --(3,-0.3) node [below] {$t^{0,0}_2$} 
(4.5,0) -- (4.5,-0.3) node [below] {$t^{0,1}_1$}
(6.5,0) --(6.5,-0.3) node [below] {$t^{0,0}_3$}
(8,0)--(8,-0.3)  node [below] {$t^{0,0}_4$}
(9.5,0) --(9.5,-0.3) node [below] {$t^{1,2}_1$}
(11,0) --(11,-0.3) node [below] {$t^{0,1}_2$}
(12.5,0) -- (12.5,-0.3) node [below] {$t^{0,0}_5$}
(14,0) -- (14,-0.3) node [below] {$t^{0,2}_1$}
(15.5,0) -- (15.5,-0.3) node [below] {$t^{0,0}_6$};
\draw (0,1) -- (-0.3,1) node [left] {\small $1$};
\draw (0,2) -- (-0.3,2) node [left] {\small $2$};
\draw (0,3) -- (-0.3,3) node [left] {\small $3$};
\draw (0,4) -- (-0.3,4) node [left] {\small $4$};
\draw [very thick,dashed] (0,0) -- ++(1,0) -- ++(0,1)  -- ++(2,0) -- ++(0,1) -- ++(1.5,0) -- ++(0,-2) 
-- ++(2,0) -- ++(0,1) -- ++(1.5,0) -- ++(0,1) -- ++(3,0) -- ++(0,-2) -- ++(1.5,0) 
-- ++(0,1) -- ++(3,0) -- ++(0,1) -- ++(0.5,0) node [right]  {$X_1(t)$};
\draw [very thick, blue, dotted] (0,0) -- ++(1,0) -- ++(0,1)  -- ++(2,0) -- ++(0,1) -- ++(3.5,0) -- 
++(0,1) -- ++(1.5,0) -- ++(0,1) -- ++(1.5,0) -- ++(0,-2) -- ++(3,0) -- ++(0,1) 
-- ++(1.5,0) -- ++(0,-3) -- ++(1.5,0) -- ++(0,1) -- ++(0.5,0) node [right] {$X_2(t)$};
\end{tikzpicture}
\caption{Fresh updates from a source pass through the network as point processes; $t^{i,j}_n$ marks the $n$th update sent on link $(i,j)$. Node $1$ gets updates  from the source node $0$. Node $2$ gets updates from both the source and also from node $1$. Age is measured in versions, $X_i(t)$ records how many versions out-of-date the update at node $i$ is relative to the source.}
\label{fig:cloudnet}
\vspace{-5mm}
\end{figure}

Status updates of a source node $0$ are shared via  a network with a set of nodes $\Ncal=\set{1,2,\ldots,n}$. Motivated by sensor networks in which accurate clocks may be unavailable,  timeliness at each node is measured by {\em update versions}. The source node $0$ maintains the current (fresh) version of its status and thus node $0$ always has age $X_0(t)=0$. Starting at time $t=0$, status updates at node $0$ occur as a rate $\lambda_{00}$ Poisson process $N_0(t)$. That is, at time $t>0$, the most recent update at node $0$ is version $N_0(t)$.   If  the current update  at node $i$ is version $N_i(t)$, then the age at node $i$, as measured in versions,   is $X_i(t)=N_0(t)-N_i(t)$. An example of version age  sample paths is depicted in Figure~\ref{fig:cloudnet}. 
In particular, if node $0$ has an update at time $t$, the age at each node $i$ becomes $X'_i(t)=X_i(t)+1$. 
On the other hand, $X_i(t)=0$ when node $i$ has observed the current update version of node $0$.  In this sense, the version AoI metric is similar to the age of incorrect information \cite{Maatouk-ton2020aoii} and  age of synchronization  \cite{Zhong-YS-isit2018} metrics.  For all three metrics, the age at a node is zero as long as that node has the current status of the source.

In this work, we develop a method for evaluating the limiting average age $\limty{t}\E{X_i(t)}$, which we refer to as the version AoI at node $i$. Building on prior work \cite{Yates-Kaul-IT2019,Yates-it2020}, this paper employs the methodology of the stochastic hybrid system (SHS) to analyze the convergence of the expected age.

Specifically, we assume the nodes forward updates using gossip. Node $i$ sends its most recent update to node $j$ as a  rate 
 $\lambda_{ij}$ Poisson process. If node $i$ sends its update to node $j$ at time $t$, the age at node $j$ becomes
 \begin{equation}\eqnlabel{ijupdate}
 X'_j(t)=\min[X_i(t),X_j(t)].
 \end{equation}
Implicit in \eqnref{ijupdate} is that updates are version-stamped so that node $j$ can adopt fresher updates from node $i$ but ignore older updates. 

The SHS approach is to  develop a set of ordinary differential equations for $\E{X_i(t)}$ that enables the evaluation of the limiting age $\limty{t}\E{X_i(t)}$. As we see in \eqnref{ijupdate}, this will require the characterization of age variables such as $X_{\set{i,j}}(t)\equiv\min(X_i(t),X_j(t))$. More generally, for arbitrary subsets $S\subseteq \Ncal$, the analysis will need to track the age
\begin{subequations}
\begin{equation}\eqnlabel{XS-defn}
X_S(t)\equiv\min_{j\in S}X_j(t)
\end{equation}
and its expected value
\begin{equation}
v_S(t)\equiv\E{X_S(t)}.
\end{equation}
\end{subequations}
We can interpret $X_S(t)$ as the status age of an observer of  updates arriving at any node in $S$ and we may refer to $X_S(t)$ as the (version) age of subset $S$.

The main result of the paper is the development of a system of linear equations for   the calculation of the limiting average age $\vbar_S=\limty{t}\E{X_S(t)}$. To describe this system of equations, define the update rate of node $i$ into set $S$ as
\begin{align}\eqnlabel{neighbor-rate}
\neighbor{i}(S)\equiv\begin{cases} 
 \sum_{j\in S}\lambda_{ij} & i\not\in S,\\
 0 & i\in S,
 \end{cases}
\end{align}
and the set of updating neighbors of $S$ as
\begin{align}\eqnlabel{neighbor-set}
N(S)\equiv\set{i\in\Ncal\colon \lambda_i(S)>0}.
\end{align}
With this notation, we state our main result.
\begin{theorem}\thmlabel{vbar:soln}
  The expected status age $v_S(t)=\E{X_S(t)}$ of an observer of node set $S$ converges to $\vbar_{S}=\limty{t} v_S(t)$ satisfying
\begin{align}\eqnlabel{vbar:soln}
\vbar_S= 
\frac{\lambda_{00}+
\sum_{i\in N(S)} \neighbor{i}(S)\vbar_{S\cup\set{i}}}{\lambda_0(S)+
\sum_{i\in N(S)}\neighbor{i}(S)}.
\end{align}
\end{theorem}
Proof of this claim is deferred to Section~\ref{sec:proof}. 

In Section~\ref{sec:applications}, we demonstrate the use of  \Thmref{vbar:soln} first for the $n=3$ node network in Figure~\ref{fig:network} 
and second for the $n$ node symmetric  gossip network on a complete graph,  as depicted in Figure~\ref{fig:complete} for $n=6$ nodes. In the complete graph, $\lambda_{ij}=\lambda/(n-1)$ for all node pairs $i,j\in \Ncal$.  This corresponds to each node $i\in \Ncal $ randomly sending its current updates to each of the other $n-1$ nodes as a rate $\lambda/(n-1)$ Poisson process. In addition, the source sends symmetrically  to each node $j\in\Ncal$ with Poisson rate $\lambda_{0j}=\lambda/n$. By exploiting the symmetry of the complete graph, \Thmref{vbar:soln} shows that the average age at a node grows as $\log n$.
\begin{theorem}\thmlabel{age-complete}
For the symmetric complete gossip network with the source sending updates to each node $i\in\Ncal$ at rate $\lambda/n$, the average version age of each node $i$ is
\begin{equation}
\frac{\lambda_{00}}{\lambda}\bracket{\frac{n-1}{n}\sum_{k=1}^{n-1} \frac{1}{k} +\frac{1}{n}}\le
\limty{t}\E{X_i(t)}\le\frac{\lambda_{00}}{\lambda}\sum_{k=1}^{n}\frac{1}{k}.
\end{equation}
\end{theorem}
Hence, as the network size $n$ grows, the average age at each node only grows logarithmically in $n$.  Although the communication models are different in various small ways, this average result is analogous to \cite[Theorem~3.1]{shah2009gossip} in  which the $\epsilon$-dissemination time, i.e. the time until the probability a source message has not reached all nodes is less than $\epsilon$, is shown to grow as $O(\log n)$. 
\section{Related Work}
\label{sec:related}
AoI analysis of updating systems started with the analyses of status age in single-source single-server first-come first-served (FCFS) queues \cite{Kaul-YG-infocom2012}, the M/M/1 last-come first-served (LCFS) queue with preemption in service \cite{Kaul-YG-ciss2012}, and the M/M/1 FCFS system with multiple sources \cite{Yates-Kaul-isit2012}.  Here we discuss AoI contributions relating to networks carrying the updates of a single source, as in this work. A more extensive overview of AoI research can be found in \cite{Yates-2020survey}. 
 
 To evaluate AoI for a single source sending updates through a network cloud \cite{Kam-KE-isit2013random} or through an M/M/$m$ server \cite{Kam-KE-isit2014diversity,Kam-KNE-IT2016diversity,Yates-isit2018}, out-of-order packet delivery was the key analytical challenge. 
The first evaluation of the average AoI over multihop network routes \cite{Talak-KM-allerton2017} employed a discrete-time version of the  status sampling network also employed in \cite{Yates-aoi2018,Yates-it2020}.  These works obtained simple AoI results because  the updates followed a single path to a destination monitor. This avoided the complexity of multiple paths and the consequent accounting for repeated  and out-of-order update message deliveries.

When multiple sources employ wireless networks subject to interference constraints, AoI has been analyzed under a variety of link scheduling strategies \cite{He-YE-IT2018,Lu-JL-mobicom2018,Talak-KM-2018distributed,Talak-KM-wiopt2018perfectCSI,Talak-KKM-isit2018,Maatouk-AE-ITW2018,Yang-AQP-globecom2019,Buyukates-SU-JCN2019,Leng-Yener-2019TCCN}.  Age bounds were developed from graph connectivity properties  \cite{Farazi-KB-JCN2019} when each node needs to update every other node. For DSRC-based vehicular networks, update piggybacking strategies were developed and evaluated \cite{Kaul-YG-globecom2011piggybacking}.

When update transmission times over network links are exponentially distributed,  sample path arguments were used \cite{Bedewy-SS-isit2016,Bedewy-SS-isit2017,Bedewy-SS-ToN2019} to show that a preemptive Last-Generated, First-Served (LGFS) policy results in smaller age processes at all nodes of the network than any other causal policy. Note that the status sampling network model in this work can also be viewed as a network of preemptive LGFS server; see \cite{Yates-it2020} for details. With that equivalence, \cite{Bedewy-SS-ToN2019} and this work can be viewed as complementary in that \cite{Bedewy-SS-ToN2019} proves the  age-optimality of LGFS policies and this work provides analytic tools for the evaluation of those policies.   



\section{Applications of \Thmref{vbar:soln}}\label{sec:applications}

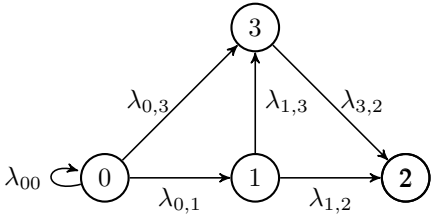
\begin{figure}[t]
\centering
\begin{tikzpicture}[->, >=stealth', auto, semithick, node distance=2cm]
\tikzstyle{every state}=[fill=none,minimum size=18pt,draw=black,thick,text=black,scale=1]
\node[state]    (0)                   {$0$};
\node[state]    (1)[right of=0]  {$1$};
\node[state]    (2)[right of=1]  {$2$};
\node[state] (3)[above of=1]  {$3$};
\node[state,scale=1] (2)[right of=1] {2};
\path
(0) edge[loop left] 	node{$\lambda_{00}$} 	(0)
(0) 	edge[below] node {$\lambda_{0,1}$} (1) 
(1) edge[right] node {$\lambda_{1,3}$}(3)
(1) edge[below] node {$\lambda_{1,2}$}(2)
(0)     edge[above,left] node {$\lambda_{0,3}$}  (3)  
(3) edge[above,right] node {$\lambda_{3,2}$}  (2);
\end{tikzpicture}
\caption{Updates generated at node $0$ are forwarded to nodes $1$, $2$, and $3$.}
\label{fig:network}
\vspace{-5mm}
\end{figure}

To utilize \Thmref{vbar:soln}, suppose we wish to calculate the average age at node $n$. We start with $S=\set{n}$ and generate an equation for $\vbar_{\set{n}}$ in terms of the variables $\vbar_{\set{i,n}}$ for nodes $i$ such that $\lambda_{i,n}>0$.  For each such node $i$, the next step is to apply \eqnref{vbar:soln} recursively  with $S=\set{i,n}$.  This generates an equation for each $\vbar_{\set{i,n}}$ in terms of variables $\vbar_{i,j,n}$ for each node $j$ that sends updates to one or both nodes in $\set{i,n}$.  

In general, at stage $k$, we construct equations for $\vbar_{S}$ for sets $S$ with size $\abs{S}=k$ in terms of variables $\vbar_{S'}$ such that each $S'$ has size $\abs{S'}=k+1$. In the worst case, this procedure terminates at stage $k=n$ when $S=\Ncal$. 
For a fully connected graph, this procedure generates equations for all $2^n-1$ non-empty subsets of $\Ncal$. On the other hand, when the network graph is sparse, substantially fewer equations may be generated.

In the next three sections, we demonstrate \Thmref{vbar:soln} with three examples; a three-node toy network with arbitrary rates, version age analysis of the  $n$-node symmetric complete graph that provides the proof of \Thmref{age-complete}, and an $n$-node symmetric ring network. 

\subsection{Toy example of \Thmref{vbar:soln}}\label{sec:simple}
Here we demonstrate \Thmref{vbar:soln} by solving for the average version age $\vbset{2}$ at node $2$ for the network shown in Figure~\ref{fig:network}.  The recursive application of \eqnref{vbar:soln} with $S=\set{2}$, $S=\set{1,2}$, $S=\set{2,3}$ and $S=\set{1,2,3}$ yields
\begin{subequations}\eqnlabel{network0132}
\begin{align}
\vbset{2}&=\frac{\lambda_{0,0}+\lambda_{1,2}\vbset{1,2}+\lambda_{3,2}\vbset{2,3}}{\lambda_{1,2}+\lambda_{3,2}},\\
\vbset{1,2}&=\frac{\lambda_{0,0}+\lambda_{3,2}\vbset{1,2,3}}{\lambda_{0,1}+\lambda_{3,2}},\\
\vbset{2,3}&=\frac{\lambda_{0,0}+(\lambda_{1,2}+\lambda_{1,3})\vbset{1,2,3}}{\lambda_{0,3}+\lambda_{1,2}+\lambda_{1,3}},\\
\vbset{1,2,3}&=\frac{\lambda_{0,0}}{\lambda_{0,1}+\lambda_{0,3}}.\eqnlabel{vbar123}
\end{align}
\end{subequations}
We note that \eqnref{vbar123} is an example of the general result that $\vbar_{\Ncal} =\lambda_{00}/\lambda_0(\Ncal)$.  For this network, it follows from \eqnref{network0132} that
\begin{align}
\vbset{2}&=\frac{\lambda_{0,0}}{\lambda_{1,2}+\lambda_{3,2}}
\left[1+
\frac{\lambda_{1,2}}{\lambda_{0,1}+\lambda_{3,2}}
\paren{1+\frac{\lambda_{3,2}}{\lambda_{0,1}+\lambda_{0,3}}}\right.
\nn
&\qquad\quad\left.+\frac{\lambda_{3,2}}{\lambda_{0,3}+\lambda_{1,2}+\lambda_{1,3}}
\paren{1+\frac{\lambda_{1,2}+\lambda_{1,3}}{\lambda_{0,1}+\lambda_{0,3}}}\right].\eqnlabel{0132example}
\end{align}
The solution \eqnref{0132example} is complicated because it includes a variety of special cases. 
For example, when $\lambda_{1,2}\to\infty$, $\vbset{1,2,3}$ and $\vbset{1,2}$ are unchanged but 
$\vbset{2}\to\vbset{1,2}$ because nodes $1$ and $2$ become equivalent to a single node with update rates $\lambda_{0,1}$ from node $0$ and $\lambda_{3,2}$ from node $3$.
On the other hand,  when $\lambda_{1,2}\to0$, $\vbset{1,2,3}$ is unchanged while
\begin{align}
\vbset{2}\to\frac{\lambda_{0,0}}{\lambda_{3,2}}+\vbset{2,3},\quad
\vbset{2,3}\to\frac{\lambda_{0,0}+\lambda_{1,3}\vbset{1,2,3}}{\lambda_{0,3}+\lambda_{1,3}}. 
\end{align}
In this case, the solution for $\vbset{2}$ reflects the path diversity offered  by the two paths from the source to node $2$.

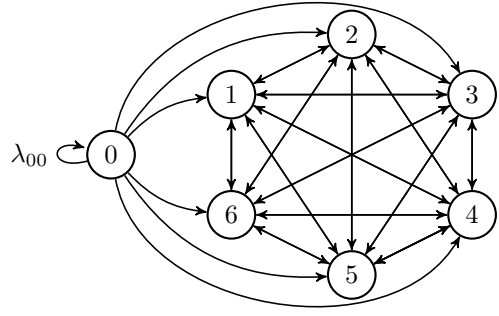
\begin{figure}[t]
\vspace{-10mm} 
\centering
\begin{tikzpicture}[->, >=stealth', auto, semithick, node distance=1cm,scale=0.8]
\tikzstyle{every state}=[fill=none,minimum size=18pt,draw=black,thick,text=black,scale=1]
\node[state]   at (-1,0) (0)                   {$0$};
\node[state]   at (1,1)  (1)  {$1$};
\node[state]   at (3,2)  (2)  {$2$};
\node[state]  at (5,1)  (3)  {$3$};
\node[state] at (5,-1) (4) {$4$};
\node[state] at (3,-2) (5) {$5$};
\node[state] at (1,-1) (6) {$6$};
\path
(0) edge[loop left] node{$\lambda_{00}$} 	(0);
\path 
(1) edge (2) (2) edge (1)
(2) edge (3) (3) edge (2)
(3) edge (4) (4) edge (3)
(4) edge (5) (5) edge (4)
(5) edge (6) (6) edge (5)
(6) edge (1) (1) edge (6);
\path 
(1) edge (3) (3) edge (1)
(1) edge (4) (4) edge (1)
(1) edge (5) (5) edge (1)
(2) edge (4) (4) edge (2)
(2) edge (5) (5) edge  (2)
(2) edge (6) (6) edge (2)
(3) edge (5) (5) edge (3)
(3) edge (6) (6) edge (3)
(4) edge (5) (5) edge (4)
(4) edge (6) (6) edge (4);
\path 
(0) edge[bend left=20] (1)
(0) edge[bend left=30]  (2)
(0) edge[bend left=70]  (3)
(0) edge[bend right=20] (6)
(0) edge[bend right=30] (5)
(0) edge[bend right=70] (4);
\end{tikzpicture}
\vspace{-5mm}
\caption{Updates generated at node $0$ at rate $\lambda_{00}$ are forwarded to nodes in $\Ncal=\set{1,\ldots,6}$ which form a complete graph. Node $0$ sends updates to each node $i\in\Ncal$  at rate $\lambda/6$. Each node $i\in\Ncal$ send updates to every other node $j$ at rate $\lambda/5$.}
\label{fig:complete}
\vspace{-5mm}
\end{figure}

\subsection{Proof of \Thmref{age-complete}}\label{sec:complete}
We now use \Thmref{vbar:soln} to find the average of a node for the$n$-node complete graph, as depicted  for $n=6$ in Figure~\ref{fig:complete}. Here the symmetry of the complete graph is essential to derive \Thmref{age-complete}. In the absence of symmetry, the recursion of \Thmref{vbar:soln} 
would generate equations for all $2^n-1$ nontrivial subsets of $\Ncal$.  

Let $S_j$ denote an arbitrary $j$-node subset of the complete graph. By symmetry, the age processes $X_{S_j}(t)$ for all subsets $S_j$ are statistically identical. Hence we define $\vtil_j=\vbar_{S_j}$. Moreover, each subset $S_j$ has $\abs{N(S_j)}=n-j$ neighbor nodes $i$ that send updates to $S_j$ at rate $\lambda_i(S_j)=j\lambda/(n-1)$. For each such neighbor $i$, $S_j\cup \set{i}$ is a $j+1$ node subset $S_{j+1}$.
Also, because the source symmetrically updates all nodes in $\Ncal$, each subset $S_j$ receives updates from the source node at rate $\lambda_0(S_j)=j\lambda/n$.  Thus  \Thmref{vbar:soln} yields
\begin{IEEEeqnarray}{rCl}
\vtil_{j}&=&\frac{\lambda_{00}+\abs{N(S_j)}\lambda_i(S_j)\vtil_{j+1}}{\lambda_0(S_j)+N(S_j)\lambda_i(S_j)}
=\frac{\lambda_{00}+\frac{j(n-j)\lambda}{n-1}\vtil_{j+1}}{\frac{j\lambda}{n} +\frac{j(n-j)\lambda}{n-1}}.\IEEEeqnarraynumspace\eqnlabel{vtil-iteration}
\end{IEEEeqnarray}
For $j=n$, $S_j=S_n$
is the set of all nodes. With all nodes  in the observer set,  the neighbor set $N(\Ncal)$ is empty, $\lambda_0(\Ncal)=\lambda$, and \Thmref{vbar:soln} yields 
$\vtil_n=\vbar_{\Ncal}=\lambda_{00}/\lambda$. With this initial condition, \eqnref{vtil-iteration} enables iterative computation of  $\vtil_{n-1},\vtil_{n-2},\ldots$ until we reach $\vtil_1$, the average age of a single node. However to complete the proof, let $j=n-k$, implying
\begin{align}
\vtil_{n-k} &=\frac{\frac{\lambda_{00}}{(n-k)\lambda}+\frac{k}{n-1}\vtil_{n-k+1}}{\frac{1}{n} +\frac{k}{n-1}}.\eqnlabel{vtil1}
\end{align}
With the definition $\vhat_k\equiv\vtil_{n-k+1}$, \eqnref{vtil1} becomes
\begin{align}
\vhat_{k+1} &=\frac{\frac{\lambda_{00}}{(n-k)\lambda}+\frac{k}{n-1}\vhat_{k}}{\frac{1}{n} +\frac{k}{n-1}}
\le \frac{\frac{\lambda_{00}}{(n-k)\lambda}+\frac{k}{n}\vhat_{k}}{\frac{1}{n} +\frac{k}{n}}.
\eqnlabel{vhat-upper}
\end{align}
The upper bound in \eqnref{vhat-upper} holds iff 
$\vhat_{k}\le n\lambda_{00}/(n-k)\lambda$.
Since $\vhat_1=\vtil_{n}=\lambda_{00}/\lambda$ this requirement holds at $k=1$ and can be shown by induction to hold for all $k$.  Defining $y_k=k\vhat_k/n$, it follows from \eqnref{vhat-upper} that
\begin{align}
y_{k+1}&\le \frac{\lambda_{00}}{(n-k)\lambda} +y_k.\eqnlabel{y_k-bound}
\end{align}
Since $y_1=\lambda_{00}/(n\lambda)$, it follows from \eqnref{y_k-bound} that
\begin{align}
y_n\le \frac{\lambda_{00}}{\lambda}\sum_{k=0}^{n-1} \frac{1}{n-k}
= \frac{\lambda_{00}}{\lambda}\sum_{k=1}^{n} \frac{1}{k}.
\end{align}
Since $y_n=\vhat_n=\vtil_1$, 
this completes the proof of the \Thmref{age-complete} upper bound. For the lower bound, the equality in \eqnref{vhat-upper} implies
\begin{align}\eqnlabel{vhat-lower}
\vhat_{k+1} &\ge\frac{n-1}{k-1}\bracket{\frac{\lambda_{00}}{(n-k)\lambda}+\frac{k}{n-1}\vhat_{k}}.
\end{align}
Defining $\yhat_k\equiv k\vhat_k/(n-1)$, \eqnref{vhat-lower} implies
\begin{align}
\yhat_{k+1}\ge \frac{\lambda_{00}}{(n-k)\lambda}+\yhat_{k}.\eqnlabel{yhat-lower}
\end{align}
It follows  from \eqnref{yhat-lower} that $\vhat_n=(n-1)\yhat_n/n$ satisfies the lower bound of \Thmref{age-complete}.
\begin{figure}[t]
\centering
\includegraphics[scale=1.2]{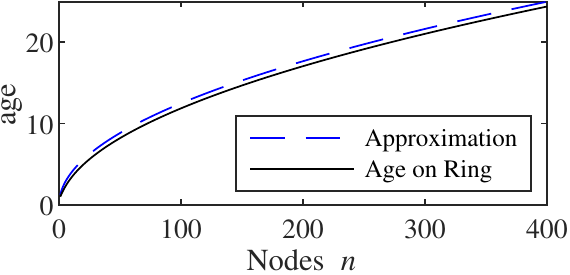}
\caption{Average version age of a node on the symmetric $n$ node ring with $\lambda_{00}/\lambda=1$. The dashed blue line $1.25\sqrt{n}$  is an empirical approximation, but not an upper bound;  the age will exceed the approximation for $n>40401$.}
\label{fig:age-ring}
\vspace{-5mm}
\end{figure} 

\subsection{Age on a Symmetric Ring}\label{sec:ring}
In the ring network, the source sends updates to each node at rate $\lambda/n$ while each node $i$ sends updates to each of its neighbor nodes $i+1$ and $i-1$\footnote{We assume node indexing  modulo the $n$ node ring, i.e., node $n+1$ refers to node $1$ and node $1-1$ refers to node $n$.} at rates $\lambda/2$  at rate $\lambda/2$. Thus the network resembles the complete graph of Figure~\ref{fig:complete}, except the interior transitions are deleted. 

For the ring graph, let $R_j=\set{i,i+1,\ldots,i+j-1}$  denote an arbitrary contiguous  $j$-node subset of the ring. By symmetry, the age processes $X_{R_j}(t)$ for all subsets $R_j$ are statistically identical. Hence we define $\vtil_j=\vbar_{R_j}$. Moreover, for $j<n-1$, each subset $R_j$ has $\abs{N(R_j)}=2$ neighbor nodes $k\in\set{i-1,j}$ that send updates to $R_j$ at rate $\lambda_{k}(R_j)=\lambda/2$. 
For each neighbor $k$, $R_j\cup \set{k}$ is a $j+1$ node subset $R_{j+1}$.
In addition, for a subset $R_{n-1}$, there is a single neighbor  $k$ who sends updates at rate $\lambda_k(R_{n-1})=\lambda$ (at rate $\lambda/2$ to neighbor nodes $k+1$ and $k-1$  that are the head and tail of $R_{n-1}$. Thus $\abs{N(R_{j})}\lambda_k(R_j)=\lambda$ for $j\in\set{1,\ldots,n-1}$. 
Also, because the source symmetrically updates all nodes in $\Ncal$, each subset $R_j$ receives updates from the source node at rate $\lambda_0(R_j)=j\lambda/n$.  Thus  \Thmref{vbar:soln} yields
\begin{IEEEeqnarray}{rCl}
\vtil_{j}&=&\frac{\lambda_{00}+\abs{N(R_j)}\lambda_k(R_j)\vtil_{j+1}}{\lambda_0(R_j)+N(R_j)\lambda_k(R_j)}
=\frac{\lambda_{00}+\lambda\vtil_{j+1}}{\frac{j\lambda}{n} +\lambda}.
\IEEEeqnarraynumspace\eqnlabel{vtil-ring-iteration}
\end{IEEEeqnarray}
For $j=n$, $R_j=R_n$
is the set of all nodes. With all nodes  in the observer set,  the neighbor set $N(\Ncal)$ is empty, $\lambda_0(\Ncal)=\lambda$, and \Thmref{vbar:soln} yields 
$\vtil_n=\vbar_{\Ncal}=\lambda/\lambda_{00}$, as it does for all graphs. With this initial condition, \eqnref{vtil-ring-iteration} enables iterative computation of  $\vtil_{n-1},\vtil_{n-2},\ldots$ until we reach $\vtil_1$, the average age of an individual node. As of this writing, the downward iteration \eqnref{vtil-ring-iteration} has not yet  yielded a simple bound for $\vtil_1$
However, as the numerical evaluation is nearly trivial, an age plot is presented in Figure~\ref{fig:age-ring} for $\lambda_{00}/\lambda=1$. From the figure, it is empirically observed that $\age_{\text{ring}}(n)\approx 1.25\sqrt{n}$. 

This numerical evidence may seem surprising since $O(n\log n)$ dissemination time has been reported for the ring graph \cite{shah2009gossip}. However, to enable age comparisons with the complete graph, the ring model in this work sends its fresh updates randomly to the ring.  By contrast, the ring graph model in \cite{shah2009gossip} assumes the source is a node on the ring and thus the dissemination time to all nodes must be $\Omega(n)$. 
If source updates were passed only to a single node on the ring, the average age would indeed grow as $O(n)$. 

\section{Stochastic Hybrid Systems for AoI Analysis}\label{sec:SHS}
In this section we use a stochastic hybrid system  (SHS) model to derive \Thmref{vbar:soln}.
While there are many SHS variations \cite{Teel-SS-2014stability}, this work follows \cite{Yates-Kaul-IT2019,Yates-it2020}, which employ 
the model and notation in \cite{Hespanha-2006modelling}.  In general, the SHS  is described by  a discrete state  $q(t)\in\Qcal=\set{0,1,\ldots,\qmax}$ that evolves as a point process,  a continuous component  $\Xv(t)
\in\R^{\nlen}$ described by a stochastic differential equation in each state $q\in\Qcal$, and a set $\Lcal$ of transition/reset maps that correspond to both changes in the discrete state and jumps in the continuous state. 

\subsection{Version AoI for gossip networks as an SHS}\label{sec:SHS-AoI} 
In this work, the operation of the gossip network is memoryless; each node $i$ sends its current update to node $j$ as a Poisson process of rate $\lambda_{ij}$. Hence, the SHS discrete state space is the trivial set $\Qcal=\set{0}$. Furthermore, because age is measured in versions, the normally continuous age state $\Xv(t)$ in fact becomes discrete in the version gossip network. That is, $\Xv(t)$ changes only when there is a transition that corresponds to an update being forwarded. In the absence of such a transition, the stochastic differential equation of the SHS is trivially $\dot{\Xv}(t)=\zerov$. 

The remaining component of the SHS model is the set $\Lcal$ of  discrete transition/reset maps. In the gossip network, $\Lcal$ corresponds to the set of directed edges $(i,j)$ over which node $i$ updates node $j$. However, because of the special role of node $0$ as the source, there are three kinds of transitions. 
First, $(i,j)=(0,0)$ corresponds to the source node generating a new version so that the version age at all other nodes $k$ increases by one. The second type of transition is given by $(0,j)$, corresponding to the source node $0$ sending the current version to node $j$, reducing the age at node $j$ to zero. In the third type, a gossiping node $i$ forwards its current update to node $j$; node $j$ accepts the update if it is a fresher than its existing version. To summarize, the set of transitions is
\begin{equation}
\Lcal=\set{(0,0)}\cup\set{(0,j)\colon j\in\Ncal}\cup\set{(i,j)\colon i,j\in\Ncal},
\end{equation}
transition $(i,j)$ occurs at rate $\lambda_{i,j}$,  and in that transition the age vector becomes 
\begin{subequations}\eqnlabel{SHStrans}
$\phi_{i,j}(\Xv)=\vec{X'_1 &\cdots & X'_n}$
 such that 
 \begin{equation}\eqnlabel{SHSij}
 X'_k=\begin{cases}
 X_k+1 & i=0,j=0,k\in\Ncal,\\
 0 & i=0,k=j\in\Ncal,\\
 \min(X_i,X_j) & i\in \Ncal, k=j\in\Ncal,\\
 X_k &\ow.
 \end{cases}
 \end{equation}
 \end{subequations}

Because of the generality and power of the SHS model, complete characterization of the $\Xv(t)$ process is often impossible. The approach in \cite{Hespanha-2006modelling} is to define test functions $\psi(q,\Xv,t)$ whose expected values $\E{\psi(q(t),\Xv(t),t)}$ are performance measures of interest that can be evaluated as functions of time; see \cite{Hespanha-2006modelling}, \cite{Hespanha-course}, and the survey \cite{Teel-SS-2014stability} for additional background.

Since the simplified SHS for the gossip network is time invariant and has a trivial discrete state, it is sufficient to employ the time invariant test functions $\psi_S(\Xv)=X_S$. These test functions yield the processes
\begin{equation}\eqnlabel{psiS:defn}
\psi_S(\Xv(t))=X_S(t),
\end{equation} 
which have expected values 
\begin{align}
\Ebig{\psi_S(\Xv(t))}&=\Ebig{X_S(t)}\equiv v_S(t).\eqnlabel{orig-vqi0}
\end{align}
The objective here is to use the SHS framework to derive a system of differential equations  
for the $v_S(t)$. To do so, the SHS mapping $\psi\to L\psi$ known as the extended generator is applied to every test function $\psi(\Xv)$. The extended generator $L\psi$ is simply the function whose expected value is the expected rate of change of the test function $\psi$.
Specifically, a test function $\psi(\Xv(t))$  has an extended generator $(L\psi)(\Xv(t))$ that  satisfies Dynkin's formula
\begin{align}
\eqnlabel{dynkins}
\deriv{}{\E{\psi(\Xv(t))}}{t}&=\E{(L\psi)(\Xv(t))}.
\end{align}
For each test function $\psi(\Xv)$, \eqnref{dynkins} yields a differential equation for $\E{\psi(\Xv(t))}$.

From \cite[Theorem~1]{Hespanha-2006modelling}, it follows from the trivial discrete state, the trivial stochastic differential equation $\dot{\Xv}(t)=\zerov$, and the time invariance of $\psi_S(\Xv)$ in \eqnref{psiS:defn} that 
the extended generator of a piecewise linear SHS is given by
\begin{align}\eqnlabel{Lpsi-defn}
(L\psi_S)(\Xv)
&= 
\sum_{(i,j)\in\Lcal} \lambda_{ij} \bracket{\psi_S(\phi_{i,j}(\Xv))-\psi_S(\Xv)}.
\end{align}
\subsection{Proof of \Thmref{vbar:soln}}\label{sec:proof}
In \eqnref{Lpsi-defn}, it follows from  \eqnref{XS-defn}, \eqnref{SHStrans}, and \eqnref{psiS:defn}  that the effect on  the test function of transition $(i,j)$ is
\begin{align}\eqnlabel{psiSij}
\psi_S(\phi_{i,j}(\Xv))=X'_S=\min_{k\in S}X'_k.
\end{align} 
Evaluation of \eqnref{psiSij} depends on the transition type $(i,j)$, as given in \eqnref{SHStrans}. In transition $(0,0)$, the source node has a version update and each node $k\in S$ becomes one more version out of date. This implies $X'_k=X_k+1$ for all $k\in\Ncal$ and thus 
\begin{align}
X'_S=\min_{k\in S} X'_k = X_S+1.
\end{align} 
For other transitions $(i,j)$, only the age $X_j$ at node $j$ is changed. Thus if $j\not\in S$, then  $X_S=\min_{k\in S}X_k$ is unchanged. However,  if $j\in S$, then
\begin{align}
X'_S=\min_{k\in S} X'_k&=\min(\min(X_i,X_j),\min_{k\in S\backslash\set{j}} X_k)\nn
&=\min_{k\in  S \cup \set{i}} X_k
=X_{S\cup\set{i}}.\eqnlabel{newX_S}
\end{align}
In addition to the common $(i,j)$ transition in which $i\in\Ncal$ is a gossiping neighbor of $j\in S$,  we note that \eqnref{newX_S} incorporates some special cases. If $i=0$, then $X'_S=X_{S\cup\set{0}}=0$ since $X_0=0$. On the other hand, if $i\in S$, then $S\cup\set{i}=S$ and $X'_S=X_S$. That is, an update sent by a node in $S$ cannot reduce the age $X_S$.  

Based on the three types of transitions, namely $(0,0)$, $(0,j)$, and $(i,j)$, we conclude that 
\begin{align}\eqnlabel{Lpsi2}
(L\psi_S)(\Xv)
&= \lambda_{00}(X_S+1-X_S)+\sum_{j\in S} \lambda_{0j}[0-X_S]\nn
&\qquad\qquad+\sum_{\substack{i>0\\ i\not\in S}}\sum_{j\in S} \lambda_{ij}\bracket{X_{S\cup\set{i}}-X_S}.
\end{align}
We note that $\Xv$, $X_S$, and $X_{S\cup\set{i}}$ in \eqnref{Lpsi2} refer to the age processes $\Xv(t)$, $X_S(t)$ and $X_{S\cup\set{i}}(t)$. With this in mind, we  take the expectation of \eqnref{Lpsi2}.   On the left side of \eqnref{Lpsi2}, $\E{(L\psi_S)(\Xv(t))}=\dot{v}_S(t)$ by Dynkin's formula \eqnref{dynkins}. On the right side, $\E{X_S(t)}=v_S(t)$ and $\E{X_{S\cup\set{i}}(t)} =v_{S\cup\set{i}}(t)$ for all $i$. These substitutions yield
\begin{align*}
\dot{v}_S(t)\!=\! \lambda_{00}\!-\!v_S(t)\sum_{j\in S}\lambda_{0j}
+\sum_{\substack{i>0\\ i\not\in S}}\sum_{j\in S} \lambda_{ij}[v_{S\cup\set{i}}(t)-v_S(t)].
\end{align*}
Employing the definitions \eqnref{neighbor-rate} and \eqnref{neighbor-set} of the update rate $\lambda_i(S)$ of node $i$ into $S$, and the neighbor set $N(S)$, we obtain
\begin{align*}
\dot{v}_S(t)\!=\!\lambda_{00} \!-\!v_S(t)\Bigl[\lambda_0(S)\!+\!\sum_{\mathclap{i\in N(S)}}\neighbor{i}(S)\Bigr]
\!+\!\sum_{\mathclap{i\in N(S)}} \neighbor{i}(S) v_{S\cup\set{i}}(t).
\end{align*}
By setting the derivatives $\dot{v}_S(t)=0$, we obtain a linear equation for the time average age $\vbar_S=\limty{t}v_S(t)$ in terms of the necessary $\vbar_{S\cup\set{i}}$.
This yields \eqnref{vbar:soln}.

\section{Conclusion}
\label{sec:conclusions}
This work has introduced  AoI analysis tools for gossip algorithms on network graphs. In \Thmref{vbar:soln} we developed a set of linear equations for the computation of average version age at any node in a gossip network described by an arbitrary graph.
While the general solution has exponential complexity in the number of nodes, we believe this unavoidably reflects the multiplicity of paths from the source to a node.  When this method is applied to the $n$ node complete graph,  it was shown  using symmetry properties that the average version age at each node grows as $\log n$. This promising result suggests that gossip networks may indeed be suitable for low latency measurement dissemination, particularly in sensor network settings.

As age analysis for gossip networks is new, considerable work remains. Since this work has examined  only the simplest network graphs,  age analysis over more complex graphs is needed. Age analysis of gossip for energy harvesting sensors would also be  another obvious area of interest. While this work employs the version age metric, we expect to see  analogous results for the traditional sawtooth age metric  that tracks the evolution of time. We also believe it may be possible to derive distributional properties of the age in a gossip network by extending the moment generating function (MGF) approach to age analysis in \cite{Yates-it2020}.  
\newpage
\begin{spacing}{1.0}
\bibliographystyle{IEEEtran}
\bibliography{AOI-2020-05,aoi_eh_summary,gossip}
\end{spacing}
\end{document}